\documentclass[prb,twocolumn,superscriptaddress,amsmath,amssymb]{revtex4-1}
\usepackage{amssymb,amsmath}
\usepackage{color}
\usepackage{graphicx}
\usepackage{dcolumn}
\usepackage{bm}
\usepackage{latexsym,epsfig}
\usepackage{hyperref}

\usepackage{marvosym}
\usepackage{color}
\usepackage[utf8x]{inputenc}

\DeclareUnicodeCharacter{2212}{-}

\begin{document}
\title{Quantum-confined charge transfer that enhances magnetic anisotropy in lanthanum M-type hexaferrites}
\author{Churna Bhandari}
\affiliation{The Ames Laboratory, U.S. Department
of Energy, Iowa State University, Ames, IA 50011, USA}

\author{Michael E. Flatt{\'e}}
\affiliation{Department of Physics and Astronomy,
University of Iowa,
Iowa City, IA 52242,
USA}
\affiliation{Department of Applied Physics, Eindhoven University of Technology, Eindhoven, The Netherlands}
\author{Durga Paudyal}
\affiliation{The Ames Laboratory, U.S. Department of Energy, Iowa State University, Ames, IA 50011, USA}
\affiliation{Departments of Electrical and Computer Engineering, and Computer Science,  Iowa State University, Ames, Iowa 50011, USA}
\begin{abstract}
Iron-based hexaferrites are critical-element-free permanent magnet components of magnetic devices.
Of particular interest is electron-doped M-type hexaferrite i.e., LaFe$_{12}$O$_{19}$ (LaM) in which extra electrons introduced by lanthanum substitution of barium/strontium play a key role
in uplifting the magnetocrystalline anisotropy. 
 We investigate the electronic structure of lanthanum hexaferrite using a \textit{localized} density functional theory which  reproduces semiconducting behavior and identifies the origin of the very large magnetocrystalline anisotropy.  Localized charge transfer from lanthanum to the iron at the crystal's $2a$ site 
produces a narrow $3d_{z^2}$ valence band  strongly locking the magnetization along the $c$ axis.
The calculated uniaxial magnetic anisotropy energies from fully self-consistent calculations are nearly double the single-shot values, and agree well with available experiments. The chemical similarity of lanthanum to other rare earths suggests that LaM can  host for other rare earths possessing non-trivial $4f$ electronic states for, \textit{e.g.,} microwave-optical quantum transduction.

\end{abstract}
\maketitle


\section*{Introduction}
Since the discovery of M-type hexaferrites in 1950, these complex oxides have been of continuing research interest for permanent magnets and magnetic memory devices\cite{KimuraARCM012,CoeyIEEE011,Pullar012,Permanent-Magnet}, however the unique interplay between charge, spin, and orbital degrees of freedom suggest broader applications, including to quantum information science. M-type hexaferrites have a chemically and thermally stable crystal structure composed of easily accessible constituent elements in nature, especially barium hexaferrite and strontium hexaferrite\cite{CoeyIEEE011}. The Gorter's type\cite{GorterPIEE57} hexaferrite has alternately aligned parallel and anti-parallel magnetic moments of ferric Fe$^{3+}$ ions with respect to the hexagonal axis, aided by superexchange interaction via oxygen resulting in a large magnetic moment (20 $\mu_B$/f.u.). The unique crystal structure leads to a huge uniaxial magnetocrystalline anisotropy (MCA) along the hexagonal axis, which assists their usage for high-frequency microwave elements (\textit{i.e.} at low applied magnetic fields). 
An increase in the magnetic anisotropy constant ($K_1$) leads directly to higher-frequency microwave operation, assuming the microwave loss remains low. 

A known approach to enhance the MCA is through replacement of the divalent barium with trivalent lanthanum 
in M-type hexaferrites\cite{LotgeringJPCS74,KupferlingJAP05,GROSSINGERJMMM07}. The increase in the magnetic anisotropy was attributed to the extra electron added\cite{LotgeringJPCS74}, with the suggestion that it leads to the formation of one Fe$^{2+}$ per formula unit and thus enhances the orbital angular momentum. The
increase of the orbital angular momentum provides, through the  spin-orbit interaction (SOC),  a stronger magnetic anisotropy.
Although the crystal structure consists of five Fe-sublattices at inequivalent crystal site locations (labelled 2a, 2b, 12k, 4f$_1$, and 4f$_2$), it is not clear where the added electron resides.
Ref.~\onlinecite{LotgeringJPCS74} suggested the Fe (2a) site as a preferred site of electron localization, but there was no direct experimental measurement to confirm the formation of Fe$^{2+}$. 
Ref.~\onlinecite{GROSSINGERJMMM07} studied the charged states of Fe ions and inferred a partially quenched orbital moment of Fe (2a) from the anomalous behavior of the hyperfine field splitting in their M{\" o}ssbauer spectroscopy measurement. A similar charge state is discussed in other M{\" o}ssbauer experiments\cite{SeifertJMMM09,KupferlingPRB06}, but they did not conclusively determine the magnitude of orbital moment which is required to confirm the Fe$^{2+}$ character.

Standard density functional theory (DFT) calculations predict a delocalized electronic state resulting from the additional electron introduced by the lanthanum\cite{KupferlingPRB06}, and thus predicts metallic behavior in disagreement with the experimentally observed semiconducting behavior. Thus it is not surprising that these calculations also yield an MCA much smaller than experimentally measured. In LaM, although Refs.~\onlinecite{ChlanPRB015,NovakEPJ05} employed the same full-potential linearized augmented plane wave (FP-LAPW) methods, they obtained very different results, including for the spin magnetic moments of Fe.  
Ref. ~\onlinecite{NovakEPJ05} could not conclusively find localization of extra electrons, which is expected in the delocalized electron state calculation. So, the discussion of MCA is irrelevant there.
Ref.~\onlinecite{ChlanPRB015} found electrons localized at Fe (2a), however their calculated $K_1$ is smaller by a factor of two compared to the experiment for both SrM and LaM. 
A clear understanding of the proper electronic structure of lanthanum M-type hexaferrite (LaM) would thus provide a clear pathway towards describing the properties of $4f$-spin containing rare-earth-doped materials, such as are now used in other semiconducting rare-earth hosts like yttrium orthovanadate and yttrium oxide.\cite{JonathamPRB018,Bartholomew2020,SerranoPRB019}



\begin{figure}[!hbtb]
 \includegraphics[scale=0.3]{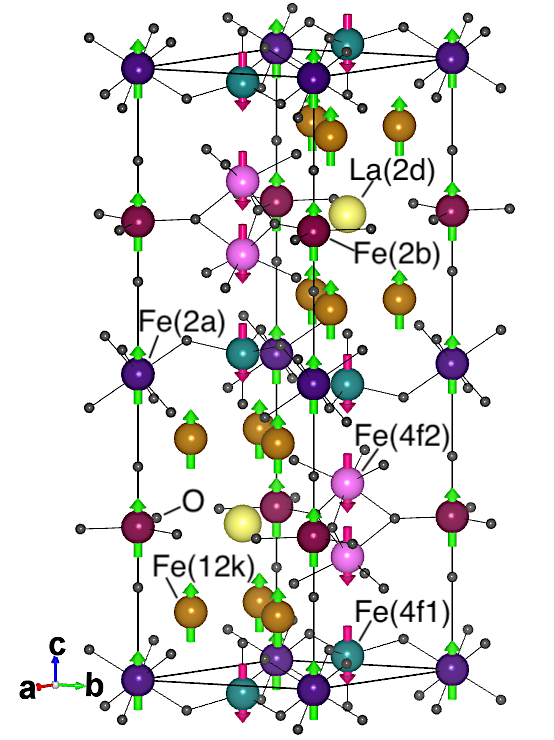}\hspace*{-0.2cm}\includegraphics[scale=0.2]{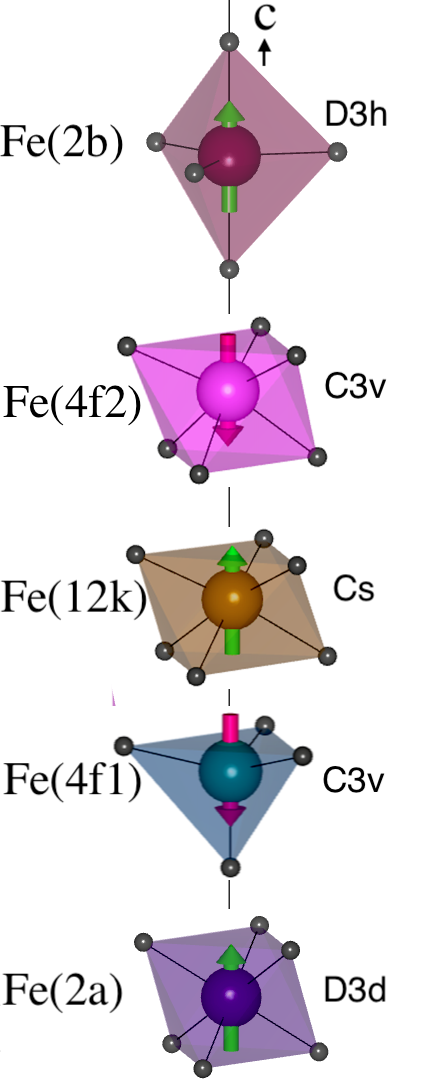}
\caption{(\textit{Left}) Crystal structure of M-type hexaferrite (LaFe$_{12}$O$_{19}$) with Gorter's-type spin configurations of the different Fe sublattices. Violet, brown, and purple balls are Fe (2a), Fe (12k), and Fe (2b) with spin-$\uparrow$(green) and dark green and magenta balls are Fe (4f$_1$) and Fe (4f$_2$) with spin-$\downarrow$ (red), respectively. Yellow and black balls are La and O atoms. (\textit{Right}) The polyhedra of each Fe-sublattice based on their nearest O atoms are shown  with the site  symmetries. \label{structure}}
\end{figure}



Here we investigate the electronic structure, optical behavior and magnetic properties of Ba/SrM and LaM using \textit{ localized} density functional theory (LDFT). The key results are: (i) replacement by La modifies the electronic band structure of BaM (SrM) significantly, producing a strongly localized band in the gap region, (ii) The electronic bandgap occurs between states of the same spin  and is reduced, (iii) La-substituted electrons are strongly localized around the Fe (2a) site and occupy a localized ($3d_{z^2})$ band, (iv) The spin magnetic moment of Fe (2a) is reduced accordingly due to the formation of an Fe$^{2+}$ charge state, and  (v) The strongly localized extra electron  enhances the $K_1$ by approximately two compared to BaM/SrM, which agrees with experiment.

\section*{Methods and crystal structure}\label{str}
The first-principles calculations were performed by using the Vienna {\it ab-initio} Simulation Package (VASP) in the all-electron (AE) projector augmented wave (PAW) form\cite{KressePRB93,KressePRB99} with the generalized gradient approximation (GGA) for the exchange-correlation functional including the Hubbard U corrections for Fe atoms. 
Following  previous work\cite{ChlanPRB015}, we used a Hubbard $U_{eff}=4.5 $ eV in our GGA + $U$ calculations within Ref.~\onlinecite{DudarevPRB98}'s simplified rotationally invariant formalism in which only the effective value of $U_{eff}=U-J$ is relevant. We used a total energy plane wave cut off of 500 eV and $7\times 7\times 1$ {\bf k}-mesh for the Brillouin zone sampling.
In VASP, the spin-orbit coupling (SOC) correction can be obtained by fully relativistic noncollinear methods\cite{KressePRB00}. To compute the magnetic anisotropy, we used both single-shot and self-consistent schemes for the total energy calculations. We did two set of DFT calculations: i) standard and ii) localized. In the localized DFT calculations a really high value of $U_{eff}$ is used and it is subsequently reduced to a realistic value to avoid a higher-energy delocalized state solution. 

We used experimental lattice parameters measured by the x-ray diffraction method\cite{KohnJPP64,KupferlingPRB06} in our calculations. We note that LaM undergoes a structural transition from hexagonal to orthorhombic at low temperature, but we neglect the low temperature structure in this work. 
The room temperature crystal structure has a hexagonal primitive unit cell with space group 194-P63/mmc (D$_{6h}^4$) as shown in Fig.~\ref{structure}, which consists of two formula units of LaFe$_{12}$O$_{19}$ and a total of 11 sublattices viz., five Fe at 2a, 2b, 12k and 4f$_1$, 4f$_2$, five O at 4f$_1$, 4e, 12k, 12k, 6h, and 1 La at 2d Wyckoff positions. Fe (2a), Fe (12k), and Fe (4f$_2$) atoms form octahedral networks with O atoms with $D_{3d}$(-3m), $C_s$(m), and $C_{3v}$(3m) point group symmetry, while Fe (4f$_1$) forms tetrahedra with $C_{3v}$(3m) and Fe (2b) forms a bipyramidal structure with $D_{3h}$(-6m2) point group symmetry as shown in Fig. \ref{structure} ({\it Right}).
We used the experimental structure parameters in our calculations to avoid the systematic lattice parameter overestimate from the GGA functional. We also did test calculations using the fully relaxed \textit{ab initio} structure and found no significant changes in the properties.

\section*{Results and discussion}\label{resul}
\begin{figure*}[!hbtp]
\hspace*{-0.25cm}\includegraphics[scale=0.4]{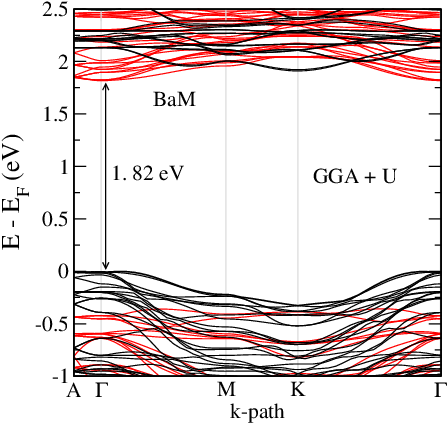}\hspace*{0.1cm}\includegraphics[scale=0.4]{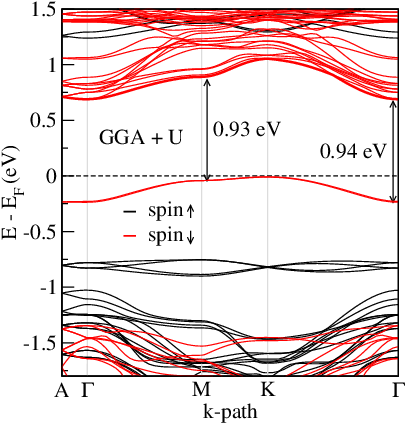}\hspace*{0.04cm}\includegraphics[scale=0.4]{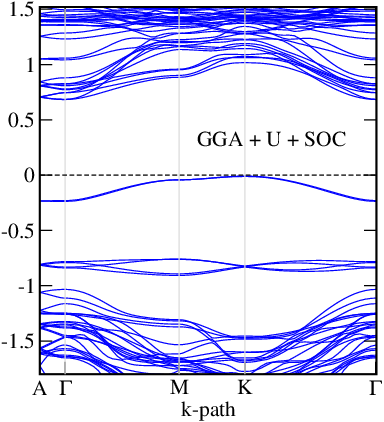}
\\\hspace*{1cm}(a)\hspace*{5cm} (b)                 \hspace*{5.5cm}      (c)
\caption{Electronic band structure calculated using GGA + U of (a) BaM and (b) LaM  and (c) GGA + U + SOC of LaM methods along  high symmetry {\bf k} points. The high symmetry points are: A = $\dfrac{1}{2}(0,0,1)$, $\Gamma = (0,0,0)$, M = $\dfrac{1}{2}(1,0,0)$, and K = $\dfrac{1}{3}(1,1,0)$ in units of primitive reciprocal lattice vectors, where $\vec{a} = a(1,0,0)$, $\vec{b} = a(-1/2, \sqrt{3}/2, 0)$, and $\vec{c} = (0, 0,c)$ are primitive translation lattice vectors.}\label{band}
\end{figure*}
\subsection*{Evolution of electronic structure with La-substitution}

We begin our discussion with the electronic band structure of the Gorter's type BaM ferrimagnet calculated along  high symmetry {\bf k} points as shown in Fig.~\ref{band}. The electronic band gap occurs between spin majority valence bands of Fe ($3d \uparrow$) and  spin minority conduction bands of Fe ($3d \downarrow)$. The valence bands consist of a strong admixture of Fe ($3d \uparrow$) with
oxygen ($2p$), whereas conduction bands are mostly Fe ($3d \downarrow)$. As is shown in Fig. \ref{band}(a), BaM has a direct band gap of 1.82 eV at $\Gamma$, computed with GGA + U methods, in excellent agreement with recent experiment\cite{AlanML019} ($\sim 1.82-1.97$ eV).
The lowest indirect band gap of 1.84 eV between the valence band maximum (VBM) (at A) and the conduction band minimum (CBM) (at $\Gamma$) is also not different from the direct gap as observed in the experiment. Although the experimentally observed indirect gap ($\sim 1.72-1.77$ eV) is slightly smaller than the direct gap, one cannot really distinguish them since they are very close. Also, the high symmetry ${\bf k}$-point A is pretty much the same as $\Gamma$ because the lattice constant $c$ is much larger ($c\sim 4 a$) as compared to $a$. 

The optical band gap related to the transitions between the valence and conduction bands with the same spins is larger by $\sim 0.5$ eV than the lowest direct gap. We found quite a bit increase of gap (2.34 eV) in the optimized structure, but the other properties such as magnetic moment and magnetic anisotropy constant remain the same as discussed elsewhere\cite{ChlanPRB015}. 
We also note that the magnitudes of calculated band gaps depend on the choice of Hubbard parameters, $U$ and $J$. The sensible value of $U_{eff}=4.5$ eV used in the calculations is based on its similarity to that obtained for Fe$^{3+}$ in $\alpha$-Fe$_2$O$_3$ with spectroscopic measurements\cite{Zimmermann99}.  

Next, we discuss the electronic band structure of LaM.  Ordinary DFT predicts a delocalized state of LaM leading to a half-metal (this band structure is shown in the Appendix Fig.~\ref{appendix_fig2}). In contrast, experimentally LaM is a semiconductor\cite{KupferlingJAP05}. 
The calculated  electron from La-substitution is spread over all the Fe-sublattices, even though the occupancy is  larger on Fe (2a) than the other iron sites. In essence, the small amount of the extra electron, when shared among the Fe sites, will not  modify the Fe$^{3+}$ charge state to Fe$^{2+}$. 
Indeed, the calculated values of the spin magnetic moments confirm these results, which are about the same ($4.10 \mu_B$) for all the, Fe atoms as found in undoped BaM.

The true ground state is obtained by the localized DFT calculation which predicts correctly LaM as a semiconductor agreeing with the experiment. The calculated total energy per unit cell is also about 1 eV smaller than that of the delocalized solution, confirming the global ground state property of the localized result.
Since the delocalized solution is metastable, hereafter, we only discuss the results from the localized calculations.

Previous calculations in Ref. ~\onlinecite{NovakEPJ05,ChlanPRB015} did not study the band structure apart from structural change and charge state. In Ref.~\onlinecite{NovakEPJ05} authors calculated the density of states but it shows an incorrect metallic state.
Therefore, to the best of our knowledge, the correct band structure of LaM has not yet been studied in the literature.

La substitution dramatically changes the band structure of BaM near the Fermi level. One can see the fully occupied two distinct subbands in the gap region in Fig. \ref{band}(b). The lower subbands consist of four and the upper subbands consist of a doubly degenerate band. 
A small indirect band gap of 0.77 eV opens up between the VBM at K and the CBM at $\Gamma$, which is about half of the BaM band gap. 
The reduction of the band gap is expected since the new bands show up in the middle of the gap. 

Whereas our calculated band gap agrees with experimental measurements, its magnitude is larger than that obtained by the resistivity measurement (0.18 eV) in the ceramic LaM\cite{KupferlingJAP05}. The  band gap measured with this type of experiment would also be smaller for BaM (See the Ref.~\onlinecite{KupferlingJAP05} for SrM).
So, it would be interesting to see if one could perform a spectroscopic measurement of the gap in a LaM single crystal.
The other interesting feature is the appearance of a pseudo-Dirac-like cone at the K-point in the lower subbands, which may lead to anomalous behavior in transport measurements in hole-doped LaM, which is worth future investigation. The GGA + U + SOC computed band structure in Fig. \ref{band}(c) does not show much difference compared to a calculation without SOC, except for a few split-off bands, which is expected because the SOC is weaker than the crystal field splittings in $3d$ elements.

\subsection*{Optical anisotropy}
\begin{figure}[hbtb!]
\includegraphics[scale=0.3]{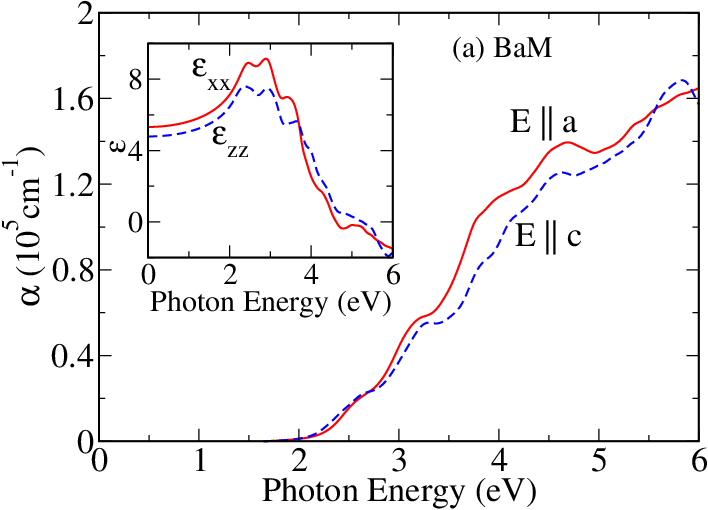}
\includegraphics[scale=0.3]{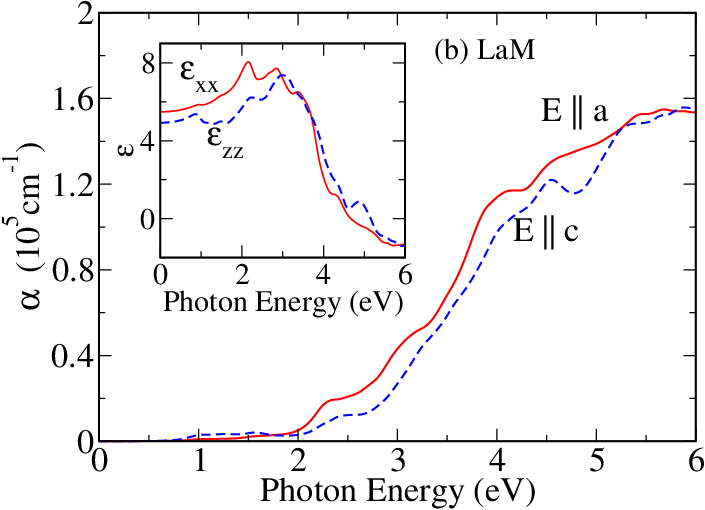}
\caption{Computed optical absorption coefficient ($\alpha$) in units of cm$^{-1}$ as a function of photon energy ( in eV) for linearly polarized light in two different directions $E \parallel a$ and $E \parallel c$ using the GGA + U method in (a) BaM and (b) LaM. Inset figures show the real part of the dielectric constant. The computed optical spectra show anisotropy between light polarization parallel to plane ($ab$) and perpendicular to plane ($c$) directions.}
\label{fig:alpha}
\end{figure}

To further scrutinize the band gap, we computed the optical absorption coefficient ($\alpha$) using  GGA + U. The results are shown in Fig.~\ref{fig:alpha}. We find $\alpha_{xx}$ for light polarization along $E\parallel a$ is stronger than $\alpha_{zz}$ for the light polarization $E\parallel c$ reflecting an anisotropy in the optical absorption in BaM. 
We note that $x$, $y$, and $z$ correspond to the crystalline $a$, $b$, and $c$ directions, respectively. The optical absorption is much weaker in the energy range (1.8 - 2.0 eV) which is expected since it corresponds to a $3d$-$3d$ transition. The optically disallowed transitions become allowed due to an admixture of Fe-$3d$ with $4s$ and $3p$ orbitals as is also seen in other $4d/5d$ oxides\cite{Bhandari2019,BhandariJPCM019}. The real part of the dielectric constants $\varepsilon_{xx}=\varepsilon_{yy}=5.32$ and  $\varepsilon_{zz}=4.79$ are consistent with the $\alpha$, which also confirms the optical anisotropy.

In LaM, the optical absorption peak is red shifted by about 1 eV as expected due to the reduced band gap. The real part of the dielectric constants,
$\varepsilon_{xx}=5.5$ and $\varepsilon_{zz}=4.95$, are similar to those in BaM. However the $\alpha$ in LaM shows a higher peak in the low energy region, differing that from BaM. 

\subsection*{Quantum-confined charge transfer}
There are five different Fe sublattices in LaM, but it is not clear which one is the most favorable for La-substituted electron localization. We address 
this issue by analyzing i) localized vs. delocalized solutions, ii) symmetry, and iii) partial density of states (PDOS).
As described above, it is now clear that the delocalized solution is not correct as it predicts LaM is a metal.
This has two main consequences: first, an electron cannot occupy the Fe atoms at 12k and 4f$_1$ sites because one electron has to be shared among multiple Fe atom that are equivalent by symmetry. This scenario would lead LaM to be a metal. Second, if an electron would occupy the 4f$_1$(4f$_2)$ sites, then it would further increase the net magnetic moment by 2 $\mu_B$ per cell, which is also not correct because the opposite magnetic moments would decrease. This contradicts with the experimentally observed magnetic moment (38 $\mu_B$ per unit cell). 

To gain a better insight into the electron localization, we employ the local symmetry of the Fe-O networks to analyze the single particle energy levels.
According to C$_{3v}$ site symmetry, the $3d$ orbitals of Fe tetrahedron at 4f$_1$ site split into $A_1$, $A_2$, and doubly degenerate $E$ irreducible representations (IRReps). In the Fe (4f$_2$) the highly distorted octahedra with C$_{3v}$ also lead to similar splittings of orbitals. The Fe (12k) octahedra has the lowest symmetry in which orbitals split into $A$ and $A^{'}$ IRReps. In either case it is inappropriate for an electron to occupy any specific empty levels because of the continuum states present due to orbital overlaps between the nearest Fe atoms. 
Then we are left with the two sublattices at the 2a and 2b sites. In this case, one can anticipate a semiconducting state if one electron occupies either one of these two sites.
\begin{figure}
 \includegraphics[scale=0.75]{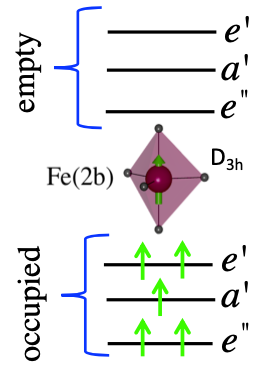}
\caption{Schematic diagram showing the crystal field splitting of single particle levels of Fe (2b) (FeO$_5$)
with $3d^5$ valence electrons. Here $a^{'}$ and $e$ are irreducible representations of the single particle $3d$ levels belonging to the $D_{3h}$ point group of Fe (2b).}\label{schematic1}
\end{figure} 
 
 \begin{figure}
 \includegraphics[scale=0.55]{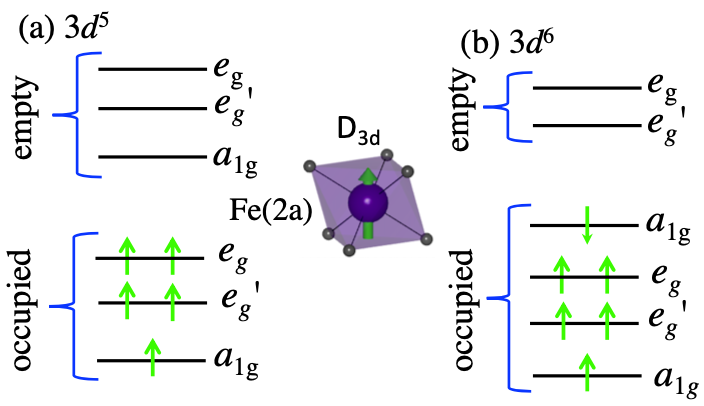}
\caption{Schematic diagrams showing the crystal field splitting of single particle levels for Fe (2a) (FeO$_6$ octahedra)
with (a) $3d^5$ and (b) $3d^6$ valence electrons. Here $a_{1g}$ and $e_g$ are irreducible representations of the single particle $3d$ levels belonging to the $D_{3d}$ point group of Fe (2a).}\label{schematic2}
\end{figure}
The local site symmetry of  Fe (2b) is D$_{3h}$ forming a bipyramidal (trigonal) FeO$_5$ unit. The Fe $(3d)$ orbitals transform as $e^{''} (dyz,dzx)$, $a^{'} (d_{z^2})$, and $e^{'} (d_{xy},d_{x^2-y^2})$ IRReps\cite{FuchikamiJPSJ065} as given in Fig.~\ref{schematic1} for the ground state configuration $A_{1}^{'}$ with S=5/2 and L=0.
The next electron would occupy the orbitally degenerate state $e^{''}$ which might lead to the well-known Jahn Teller distortion\cite{JahnTeller037}. It is inappropriate for electron to occupy this site as it would cost  extra energy.
\begin{figure*}[!htbp] 
\includegraphics[scale=0.44]{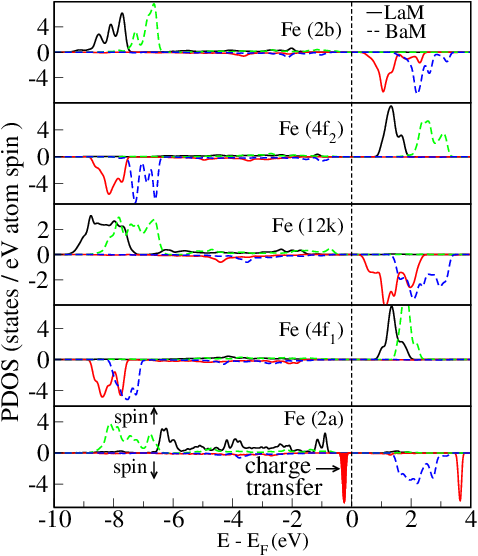}\hspace{1.75cm}\includegraphics[scale=0.55]{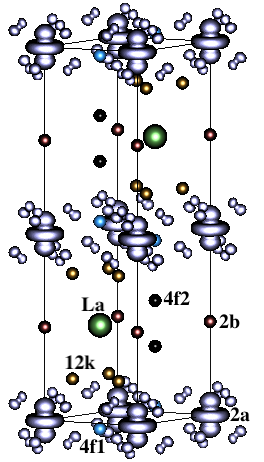}
\\\hspace*{1cm}           (a)\hspace*{5.2cm}   (b)
\caption{(a) Partial density of states (PDOS) of Fe atoms computed using GGA + U for BaM and  LaM. The shared red region below the Fermi level indicates the electron transfer to Fe (2a) from La. The solid (dashed) lines represent the PDOS in LaM(BaM) respectively. (b)
Electron charge density contours computed in occupied region of valence bands (spin-down channel) with isosurface level 0.002 e/\AA$^3$ in LaM, where Fe atoms are labeled according to their site symmetries. The contours further confirm that La-substituted extra electron charge is localized to Fe (2a) and nearby O atoms. The shape of contours around Fe atom is $d_{z^2}$-like and directed along the hexagonal axis.}\label{dos}
\end{figure*}

On the other hand, the local site symmetry of Fe (2a) is D$_{3d}$, in which Fe (3d) orbitals transform as $a_{1g}$, $e_g^{''}$, and $e_g$ IRReps. These two orbitally degenerate states with $e_g$ IRReps are two dimensional and can mix with each other while $a_{1g}$ cannot. Because of the trigonally distorted octahedral structure of Fe (2a), the ground state configuration consists of singly occupied five $3d$-orbitals in BaM and the corresponding levels are shown in Fig.~\ref{schematic2}(a). The lowest level is a $d_{z^2}$ orbital which transforms as $a_{1g}$. IRReps with respect to the hexagonal axis indicating that any extra electron would occupy it (the opposite spin channel) first as shown in Fig.~\ref{schematic2}(b). We note that the symmetry of the single particle wavefunctions does not change due to the SOC, except the splitting of Fe (12k) into two sublattices.

\begin{figure}[!htbp] 
\includegraphics[scale=0.425]{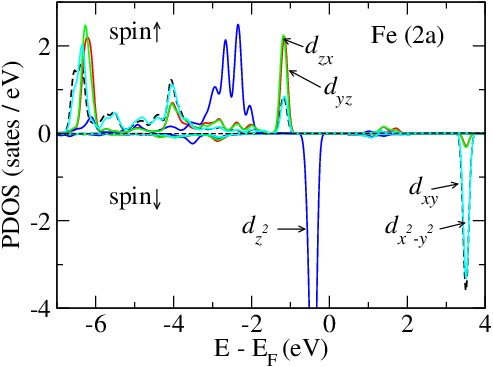}
\caption{Orbital resolved partial density of sates of Fe (2a) computed with GGA + U method in LaFe$_{12}$O$_{19}$. The sharply localized state below the Fermi level indicates the transfer of the La substituted electron into the $d_{z^2}$-derived orbital. }\label{pdos}
\end{figure}

To further confirm the electron localization as dictated by symmetry, we computed the PDOS for each atom with and without La substitution as shown in Fig.~\ref{dos}. As expected, the $5d$ states of La are located far above the Fermi level which is reflected in PDOS (shown in Appendix Fig.~\ref{appendix_fig1}) indicating that it simply acts as an electron donor (similar to Ba except it donates an extra electron).
Next, we discuss the PDOS of Fe atoms with and without La substitution. 
It is evident from Fig.~\ref{dos}(a) that the PDOS of all Fe atoms except Fe (2a) are unchanged with La substitution.
No occupied states of Fe atoms are seen near the Fermi level except for Fe (2a). But the entire $3d$-band of Fe (2a) is shifted upwards showing a sharp peak below the Fermi level. To the best of our knowledge, this is the first  prediction of such a  quantum-confined charge transfer state in the La-substituted hexaferrites. Remarkably, it lies in the gap region and it is sharply localized at the Fe (2a) site consistent with the appearance of two narrow bands (each coming from one Fe (2a)). These results suggest that the La-substituted electron occupies a single orbital of the Fe (2a) site. As shown in Fig. \ref{schematic2}, the transferred charge will occupy the $a_{1g}$ ($d_{z^2}$) orbital which is parallel to the hexagonal axis (not the  octahedral $z$-axis). To further confirm this, we computed the orbitally resolved partial density of states (Fig.~\ref{pdos}), which shows that the band is derived from the $d_{z^2}$ orbital of Fe (2a) consistent with the symmetry analysis. Electron charge distribution associated with this band is shown in Fig.~\ref{dos} (b) by a charge density contour computed in the energy range from 0.5 eV below to the Fermi level. The contour shows a perfect $d_{z^2}$ like shape pointing along the hexagonal axis at Fe (2a).  
 
To quantify the charge localization, we computed the net charge transfer to each Fe atom by integrating the occupied region of the PDOS near the Fermi level as given in Table \ref{tab1}. We find that 81\% of the added electrons occupies Fe (2a), and the remaining amount is mostly shared between O and La atoms. This in turn suggests that the extra charge added to Fe (2a) would modify its Fe$^{3+}$ charge state to Fe$^{2+}$ consistent with the anomalous behavior of the hyperfine field splitting in M{\" o}ssbauer spectroscopy\cite{SeifertJMMM09,KupferlingPRB06}.
\begin{table}
\caption{Net amount of electron transfer to different Fe atoms in LaFe$_{12}$O$_{19}$. The tabulated values are per Fe atom and are expressed in units of electrons.}\label{tab1}
\begin{ruledtabular}
\begin{tabular}{c c c  c c c c}
 Atom   &Fe (2a) &Fe (2a) &F e(12k)& F e(4f$_1$) & Fe (4f$_2$) &La/O\\
\hline
electron  &0.8127  &0.0005 &0.0038 &0.0002&0.0065 &0.1505 \\
\end{tabular}
\end{ruledtabular}
\end{table}

\subsection*{Magnetic moments and Curie temperature}
The spin magnetic moments of individual atoms in BaM/SrM and LaM are given in Table \ref{tab2}. The net magnetic moment of BaM/SrM is $39.98 \mu_B$ per unit cell which is consistent with the Gorter's type magnetic order as discussed above. 
The sum of the magnetic moments of the individual atoms is slightly smaller than the total value
as expected since these are calculated within the atomic spheres, which exclude the interstitial contributions. 
Our calculated magnetic moments for individual Fe atoms are slightly larger than experimental values due to the following factors: (1) 
 theoretical values depend on the choice of Hubbard parameters and the size of the atomic radii used in the calculations, (2)   the experimental values  measured at finite temperature  are about 10\% smaller than ideal values (5 $\mu_B$) at zero K, and (3) there is strong hybridization of Fe ($3d$) with O($p$). Indeed, the O atoms carry a spin magnetic moment as large as 0.33 $\mu_B$ as given in Table~\ref{tab2}.
\begin{table}[!htbp]
\caption{Spin magnetic moment of individual atoms in Sr/Ba/LaFe$_{12}$O$_{19}$ (Sr/Ba/LaM),  the total magnetic moment per unit cell (which includes the interstitial contribution), and comparison with experiment. Also shown are experimental magnetic moments (average values  for LaM measured at room temperature).}\label{tab2}
\begin{ruledtabular}
\begin{tabular}{l l l l l}
Atom& SrM &BaM &LaM&LaM\\
(site)& This work & This work & This work &  Expt.\\
\hline
Fe (2a)    &~4.11   &~4.23   &~3.69     &~3.35\\
Fe (2b)    &~4.08   &~4.14   &~4.15     &~3.83\\
Fe (12k)   &~4.16   &~4.24   &~4.24     &~3.73\\
Fe (4f1)   &-3.97   &-4.12   &-4.19     &-3.83  \\
Fe (4f2)   &-4.11   &-4.18   &-4.11     &-3.90\\
O (4e)     &~0.34   &~0.33   &~0.33     &\\
O (4f)     &~0.09   &~0.08   &~0.11     &\\
O (6h)     &~0.05   &~0.04   &~0.04     &\\
O (12k)    &~0.17   &~0.17   &~0.12     &\\
O (12k)    &~0.09   &~0.03   &~0.05     &\\
Sr/La (2d) &~0.00   &~0.00   &-0.01     &\\
Total     &~39.98  &~39.97  &~38.00    &\\
\end{tabular}
\end{ruledtabular}
\end{table}

All experiments see a smaller value for Fe (2a) than the other Fe sites. We again emphasize that the discrepancies between theoretical and experimental values appear to be similar for La substituted cases or not, as discussed above. 

As the calculated spin magnetic moment suggests the formation of a Fe$^{2+}$ state at the 2a site, the orbital magnetic moment expected be large following  Hund's rule. However, its magnitude is much smaller ($\sim 0.1 \mu_B$) due to the quenching of the orbital moment by the crystal  field.
The PDOS shows that 
the extra charge occupies the $d_{z^2}$ orbital with $L\sim 0$. Although the orbital moment is small, it is still increased by about 10\% in LaM as compared to SrM for Fe (2a) and it qualitatively supports the hyperfine splitting data in which the Fe at 2a site shows a transformation from Fe$^{3+}$ to Fe$^{2+}$\cite{GROSSINGERJMMM07}.

We now discuss the Curie temperature (T$_\textrm{C}$) and compare with previous results\cite{Wu2016,NovakPRB05}. We consider an isotropic exchange interaction ($J_{ij}$) between the nearest neighbor Fe atoms  within the mean field approximation. Although more involved approaches such as the random phase approximation (RPA) and mean-field approximation (MFA) have been employed previously\cite{Wu2016,NovakPRB05} to estimate T$_\textrm{C}$, its magnitude is overestimated relative to experiments when  reasonable values of the Hubbard U are used. Our straightforward calculations, 
taking the the total energy difference between ferrimagnetic and ferromagnetic configurations, show that the energy required to break a bond between the nearest neighbor Fe atoms is: $J_{ij}= 12.32$, 12.13, and 11.73 meV in SrM, BaM, and LaM, respectively.  
The total energy difference (also the energy required to break a bond) reduces with an increasing value of U thereby reducing the T$_\textrm{C}$. Remarkably, by using T$_\textrm{C}\sim S^2J_{ij}/k_B$, with $S=5/2$ for each Fe atom, we get 893, 880, and 851 K for respective compounds obtained by using a realistic value of U= 4.5 eV; these agree very well with  experimental values as shown in Table~\ref{tabCurie}. Although the magnitudes are slightly overestimated, the trend of the relative differences agrees with the experiments.

\begin{table}
\caption{Calculated Curie temperature T$_\textrm{C}$ and its comparison with previous theory and experiment for Sr/Ba/LaFe$_{12}$O$_{19}$ (Sr/Ba/LaM).}\label{tabCurie}
\begin{ruledtabular}
\begin{tabular}{l l l l}
 Methods& SrM &BaM & LaM\\
\hline
Expt. &737\cite{ZiJMM08} &723 & 695\\
This work\footnote{Hubbard, $U_{eff}$=4.5 eV} &893 &880 &851\\
Theory\footnote{Mean field approximation (MFA), $U_{eff}$=3.4 eV\cite{Wu2016}} &-&1514 & 1419\\
Theory\footnote{Random phase approximation (RPA), $U_{eff}$=3.4 eV\cite{Wu2016}} &-  &1009 & 946\\
Theory\footnote{MFA, $U_{eff}$=4 eV\cite{NovakPRB05}} &- &$\sim1500$ &-\\
Theory\footnote{RPA, $U_{eff}$=4 eV\cite{NovakPRB05}} &- &$\sim1000$ &-\\
\end{tabular}
\end{ruledtabular}
\end{table}

\subsection*{Magnetic anisotropy}
 The magnetic energy cost to rotate the spontaneous magnetization $M= M_s (\sin\theta\cos\phi \hat{x} + \sin\theta\sin\phi \hat{y} + \cos\theta \hat{z})$ with respect to the crystalline axis is the magnetic anisotropy energy ($E_a$). Here, $\theta$ and $\phi$ are magnetization angles and $\hat{x}$, $\hat{y}$, and $\hat{z}$ are unit vectors.
 Phenomenologically, the magnetic anisotropy energy density for a given magnet is given by $\dfrac{E_a}{V} = K_1 \sin^2\theta + K_2 \sin^4\theta + K_3\sin^6\theta$, where $K_i$ is the ith order magnetic anisotropy constant and $V$ is the volume of the magnet\cite{Ralph}.

\begin{table}
\caption{Calculated magnetic anisotropy constant $K_1$ in Sr/Ba/LaFe$_{12}$O$_{19}$ (Sr/Ba/LaM) and its comparison with experiment and previous DFT. The values are presented in units of MJ/m$^3$. The values outside(inside) the parenthesis correspond to results obtained by fully self-consistent (single-shot) spin-orbit coupling via noncolliiear relativistic methods.}\label{tab4}
\begin{ruledtabular}
\begin{tabular}{l l l l}
$K_1$ & SrM &BaM & LaM\\
\hline
This work\footnote{Single-shot calculations} &0.18 & 0.18 & 0.32\\
This work\footnote{Fully self-consistent calculations} & 0.34 &0.33 & 0.66\\
Expt.\cite{JahnPSS69,ShirkJAP69} & 0.35-0.36 & 0.32-0.33  &0.5-0.8  \\
Theory \footnote{FLAPW methods with U$_{\text eff}$=4.5 eV\cite{ChlanPRB015}} & 0.18 & &0.36\\ 
\end{tabular}
\end{ruledtabular}
\end{table}

For hard magnets the lowest order magnetic anisotropy constant $K_1$ dominates and its magnitude depends on the single ion and the dipolar moment (pair contribution) interaction energies. The single ion energy depends on the strength of the spin-orbit coupling ($\lambda$) and crystal  field ($\Delta$) and is much larger than the pair contribution\cite{Ralph}. 
Ref.~\onlinecite{FuchikamiJPSJ065} also showed from a model based on quantum many-body wavefunctions that 
the leading contribution is from a single ion for Fe (2b).  
The zero-field splitting (uniaxial splitting) parameter $D\sim 2$cm$^{-1}$ is two orders of magnitude larger than the dipolar spin-spin contribution. Therefore, here, we focus only on the single ion contribution to  $K_1$.

First we computed the total energy along the two different crystalline axes viz., $E_{100}$ along the $a$ and $E_{001}$ along the $c$-axis respectively. Then, the magnetic anisotropy energy is obtained as: $E_a = E_{100}-E_{001}$. 
The computed values of $K_1=E_a/V$ are positive in all cases (Table~\ref{tab4}) suggesting magnetocrystalline anisotropy with an easy axis along the crystalline $c$-axis. The values outside (inside) the parentheses are obtained with and without full self-consistent total energy calculations using the GGA + U + SOC methods.
The magnitudes of $K_1$ are sensitive to the methods and depend on whether it is a single shot or fully self-consistent calculation, despite $\lambda$'s relative weakness for Fe ($3d$) atoms. 
For instance, the computed value of $K_1$ is 0.34 MJ/m$^3$ for BaM with the fully self-consistent calculations, twice the single-shot value of 0.18 MJ/m$^3$. 
In principle the self-consistent method is a better approach because the effect of SOC is included properly in the charge density and therefore it is expected to predict better results. Indeed, the value of $K_1$ obtained with the fully self-consistent calculations is in excellent agreement with the experimental value (0.35 MJ/m$^3$). 
Similar values are predicted for $K_1$ in SrM in excellent agreement with the experiment, which is expected since in both cases we have  similar chemical and structural environments e.g., $sp$-electron elements (Sr or Ba). 

Again, in LaM the self-consistent value of $K_1$ is enhanced by about a factor of two  compared to the single-shot value, as was found for the parent compounds (Ba/SrM). The magnitude of $K_1$ is increased in LaM by a factor $\sim 2$
 compared to Ba/SrM which is in very good agreement with experiment (0.5-0.8 MJ/m$^{3}$)\cite{LotgeringJPCS74}. The increase in magnetic anisotropy can be understood in terms of charge transfer, spin, and orbital moment. 
As discussed above, the charge from La substitution occupies the $d_{z^2}$ orbital of Fe (2a) resulting in the net charge polarization along the $c$-axis along with the formation of a Fe$^{2+}$ charge state. Although the orbital moment is expected to be quenched for Fe$^{2+}$, effectively it is still increased by $\sim 10\%$ in LaM. Interestingly, it is about three times larger along the $c$-axis than in the plane, which indicates that it would cost more magnetic energy to rotate strongly polarized charge away from the $c$-axis leading to the enhancement of $K_1$ in LaM.    

\section*{Conclusion}\label{concl}
We used {\it ab-initio} calculations to study the evolution of the electronic band structure and optical spectra of the hexaferrites after  chemical substitution of a divalent $sp$-element (Ba/Sr) with a trivalent $5d$-element (La).
Our localized DFT calculations predict a very sharply localized band in the gap in LaM which is revealed by the electronic band structure and PDOS.  The origin of this localized band is due to  quantum-confined charge transfer from the substituted La to a specific site [Fe (2a)], occupying the symmetry-allowed $3d_{z^2}$-derived orbital.
The reduced value of the spin magnetic moment of Fe (2a) is consistent with localization of the additional electron there, leading to the formation of a Fe$^{2+}$ charge state at the 2a site. The increase in electronic charge, which leads to a reduction in the spin moment and an increase in the orbital moment of Fe (2a) driven by this La substitution, results in an increase of the magnetic anisotropy by a factor of two in LaM as compared to Sr/BaM and agrees very well with available experiment. Our study opens up the possibility of exploring  rare-earth-based electron-substituted hexaferrites as quantum spin systems because of the formation of this spatially localized quantum state in the gap. It also provides the description of a potential host for other $4f$ elements that would provide sharp optical transitions within the optical gap of the host, with these rare earth elements substituting for the lanthanum.

\acknowledgements{This work is supported by the Critical Materials Institute, an Energy Innovation Hub funded by the U.S. Department of Energy, Office of Energy Efficiency and Renewable Energy, Advanced Manufacturing Office. The Ames Laboratory is operated for the U.S. Department of Energy by Iowa State University of Science and Technology under Contract No. DE-AC02-07CH11358. MEF is supported by NSF DMR-1921877. We would like to thank C. \c{S}ahin for his useful remarks on the paper. We would also like to acknowledge Ed Moxley for maintaining and updating computational facilities and software.

The U.S. government retains and the publisher, by accepting the article for publication, acknowledges that the U.S. government retains a nonexclusive, paid-up, irrevocable, worldwide license to publish or reproduce the published form of this manuscript or allow others to do so, for the U.S. government purposes. U.S. Department of Energy (DOE) will provide public access to these results of federally sponsored research in accordance with the DOE Public Access Plan (http://energy.gov/downloads/doe-public-access-plan).}\\

\appendix
\section{Band structure of LaM with a delocalized electron}\label{appendix1}
\begin{figure}
\includegraphics[scale=0.5]{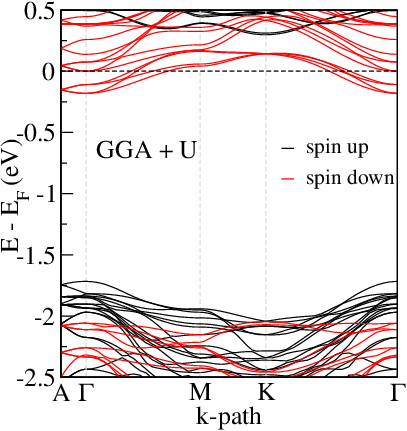}
\caption{Electronic band structure of metalstable ferrimagnet LaM calculated using the GGA + U method along  high symmetry {\bf k} points.}\label{appendix_fig2}
\end{figure}
Here we present the band structure of the ferrimagnetic metallic metastable state of LaM  shown in Fig.~\ref{appendix_fig2}, computed using delocalized (standard) DFT. In this case, the electrons from La substitution partially occupy  all the  Fe sublattices, resulting in a half metal (as visible in the figure). 

\section{Density of states of LaM}\label{appendix2}
\begin{figure}
\includegraphics[scale=0.45]{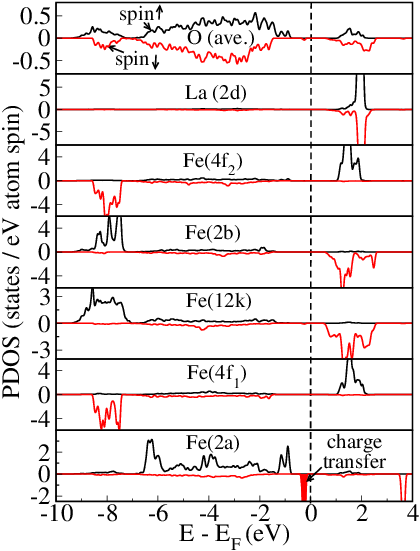}
\caption{PDOS of individual atoms in LaM where O (ave.) denotes the average PDOS of oxygens present in the unit cell.}\label{appendix_fig1}
\end{figure}

We compare the PDOS of individual atoms in LaM  in Fig.~\ref{appendix_fig1}. The oxygen PDOS shows the average value of all oxygen atoms in unit cell as denoted by O (ave.).
The La PDOS shows that $3d$ states are located farther away from the Fermi level, suggesting that it will donate an extra electron to the system. Indeed, the electron transfers to Fe (2a), because it has an extra localized band (with opposite spin) just below the Fermi level as discussed in the main text.  

\bibliography{hexaM}

\begin{thebibliography}{10}

\bibitem{KimuraARCM012}
Kimura Tsuyoshi.
\newblock Magnetoelectric hexaferrites.
\newblock {\em Annual Review of Condensed Matter Physics}, 3(1):93--110, 2012.

\bibitem{CoeyIEEE011}
J.~M.~D. {Coey}.
\newblock Hard magnetic materials: A perspective.
\newblock {\em IEEE Transactions on Magnetics}, 47(12):4671--4681, 2011.

\bibitem{Pullar012}
Robert~C. Pullar.
\newblock Hexagonal ferrites: A review of the synthesis, properties and
  applications of hexaferrite ceramics.
\newblock {\em Progress in Materials Science}, 57(7):1191--1334, 2012.

\bibitem{Permanent-Magnet}
Global permanent magnets market share report, 2020-2027.

\bibitem{GorterPIEE57}
E.~W. {Gorter}.
\newblock Saturation magnetization of some ferrimagnetic oxides with hexagonal
  crystal structures.
\newblock {\em Proceedings of the IEE - Part B: Radio and Electronic
  Engineering}, 104(5):255--260, 1957.

\bibitem{LotgeringJPCS74}
F.K. Lotgering.
\newblock Magnetic anisotropy and saturation of $\text{LaFe}_{12}\text{O}_{19}$
  and some related compounds.
\newblock {\em Journal of Physics and Chemistry of Solids}, 35(12):1633 --
  1639, 1974.

\bibitem{KupferlingJAP05}
M.~K{\"u}pferling, P.~Nov{\'a}k, K.~Kn{\` i}{\v{z}}ek, M.~W. Pieper,
  R.~Gr{\"o}ssinger, G.~Wiesinger, and M.~Reissner.
\newblock Magnetism in \text{La} substituted \text{Sr} hexaferrite.
\newblock {\em Journal of Applied Physics}, 97(10):10F309, 2005.

\bibitem{GROSSINGERJMMM07}
R.~Gr{\"o}ssinger, M.~K{\" u}pferling, M.~Haas, H.~M{\" u}ller, G.~Wiesinger,
  and C.~Ritter.
\newblock Magnetic anisotropy and magnetostriction of
  $\text{LaFe}_{12}\text{O}_{19}$.
\newblock {\em Journal of Magnetism and Magnetic Materials}, 310(2, Part
  3):2587 -- 2589, 2007.
\newblock Proceedings of the 17th International Conference on Magnetism.

\bibitem{SeifertJMMM09}
D.~Seifert, J.~T{\'o}pfer, F.~Langenhorst, J.-M.~Le Breton, H.~Chiron, and
  L.~Lechevallier.
\newblock Synthesis and magnetic properties of \text{La}-substituted m-type
  \text{Sr} hexaferrites.
\newblock {\em Journal of Magnetism and Magnetic Materials}, 321(24):4045 --
  4051, 2009.

\bibitem{KupferlingPRB06}
Michaela K\"upferling, Roland Gr\"ossinger, Martin~W. Pieper, G\"unter
  Wiesinger, Herwig Michor, Clemens Ritter, and Frank Kubel.
\newblock Structural phase transition and magnetic anisotropy of la-substituted
  $m$-type \text{Sr} hexaferrite.
\newblock {\em Phys. Rev. B}, 73:144408, Apr 2006.

\bibitem{ChlanPRB015}
Vojt\ifmmode \check{e}\else~\v{e}\fi{}ch Chlan, Karel
  Kou\ifmmode~\check{r}\else \v{r}\fi{}il, Kate\ifmmode
  \check{r}\else~\v{r}\fi{}ina Uli\ifmmode~\check{c}\else \v{c}\fi{}n\'a,
  Helena \ifmmode \check{S}\else \v{S}\fi{}t\ifmmode~\check{e}\else
  \v{e}\fi{}p\'ankov\'a, J\"org T\"opfer, and Daniela Seifert.
\newblock Charge localization and magnetocrystalline anisotropy in \text{La},
  \text{Pr}, and \text{Nd} substituted \text{Sr} hexaferrites.
\newblock {\em Phys. Rev. B}, 92:125125, Sep 2015.

\bibitem{NovakEPJ05}
P.~Nov{\'a}k, K.~Kn{\'i}\v{z}ek, M.~K{\"u}pferling, R.~Gr{\"o}ssinger, and
  M.~W. Pieper.
\newblock Magnetism of mixed valence (\text{LaSr}) hexaferrites.
\newblock {\em The European Physical Journal B - Condensed Matter and Complex
  Systems}, 43(4):509--515, Feb 2005.

\bibitem{JonathamPRB018}
Jonathan~M. Kindem, John~G. Bartholomew, Philip J.~T. Woodburn, Tian Zhong,
  Ioana Craiciu, Rufus~L. Cone, Charles~W. Thiel, and Andrei Faraon.
\newblock Characterization of $^{171}\mathrm{Yb}^{3+}:\mathrm{YVO}_{4}$ for
  photonic quantum technologies.
\newblock {\em Phys. Rev. B}, 98:024404, Jul 2018.

\bibitem{Bartholomew2020}
John~G. Bartholomew, Jake Rochman, Tian Xie, Jonathan~M. Kindem, Andrei Ruskuc,
  Ioana Craiciu, Mi~Lei, and Andrei Faraon.
\newblock On-chip coherent microwave-to-optical transduction mediated by
  ytterbium in $\mathrm{YVO}_4$.
\newblock {\em Nature Communications}, 11(1):3266, Jun 2020.

\bibitem{SerranoPRB019}
D.~Serrano, C.~Deshmukh, S.~Liu, A.~Tallaire, A.~Ferrier, H.~de~Riedmatten, and
  P.~Goldner.
\newblock Coherent optical and spin spectroscopy of nanoscale
  $\mathrm{Pr}^{3+}:\phantom{\rule{0.28em}{0ex}}\text{Y}_{2}\mathrm{O}_{3}$.
\newblock {\em Phys. Rev. B}, 100:144304, Oct 2019.

\bibitem{KressePRB93}
G.~Kresse and J.~Hafner.
\newblock Ab initio molecular dynamics for liquid metals.
\newblock {\em Phys. Rev. B}, 47:558--561, Jan 1993.

\bibitem{KressePRB99}
G.~Kresse and D.~Joubert.
\newblock From ultrasoft pseudopotentials to the projector augmented-wave
  method.
\newblock {\em Phys. Rev. B}, 59:1758--1775, Jan 1999.

\bibitem{DudarevPRB98}
S.~L. Dudarev, G.~A. Botton, S.~Y. Savrasov, C.~J. Humphreys, and A.~P. Sutton.
\newblock Electron-energy-loss spectra and the structural stability of nickel
  oxide: An \text{LSDA+U} study.
\newblock {\em Phys. Rev. B}, 57:1505--1509, Jan 1998.

\bibitem{KressePRB00}
D.~Hobbs, G.~Kresse, and J.~Hafner.
\newblock Fully unconstrained noncollinear magnetism within the projector
  augmented-wave method.
\newblock {\em Phys. Rev. B}, 62:11556--11570, Nov 2000.

\bibitem{KohnJPP64}
J.~A. Kohn and D.~W. Eckart.
\newblock New hexagonal ferrite, establishing a second structural series.
\newblock {\em Journal of Applied Physics}, 35(3):968--969, 1964.

\bibitem{AlanML019}
Alan Bañuelos-Frías, Gerardo Martínez-Guajardo, Leo Alvarado-Perea, Lázaro
  Canizalez-Dávalos, Facundo Ruiz, and Claudia Valero-Luna.
\newblock Light absorption properties of mesoporous barium hexaferrite,
  $\text{BaFe}_{12}\text{O}_{19}$.
\newblock {\em Materials Letters}, 252:239--243, 2019.

\bibitem{Zimmermann99}
R~Zimmermann, P~Steiner, R~Claessen, F~Reinert, S~H$\ddot{\text{u}}$fner,
  P~Blaha, and P~Dufek.
\newblock Electronic structure of 3d-transition-metal oxides: on-site coulomb
  repulsion versus covalency.
\newblock {\em Journal of Physics: Condensed Matter}, 11(7):1657--1682, jan
  1999.

\bibitem{Bhandari2019}
Churna Bhandari, Zoran~S Popovi{\'{c}}, and S~Satpathy.
\newblock Electronic structure and optical properties of
  $\text{Sr}_2\text{IrO}_4$ under epitaxial strain.
\newblock {\em New Journal of Physics}, 21(1):013036, Jan 2019.

\bibitem{BhandariJPCM019}
Churna Bhandari and S.~Satpathy.
\newblock Effect of epitaxial strain on the optical properties of
  $\text{NaOsO}_3$.
\newblock {\em Journal of Physics and Chemistry of Solids}, 128:265--269, 2019.
\newblock Spin-Orbit Coupled Materials.

\bibitem{FuchikamiJPSJ065}
Nobuko Fuchikami.
\newblock Magnetic anisotropy of magnetoplumbite
  $\text{BaFe}_{12}\text{O}_{19}$.
\newblock {\em Journal of the Physical Society of Japan}, 20(5):760--769, 1965.

\bibitem{JahnTeller037}
H.~A. Jahn, E.~Teller, and Frederick~George Donnan.
\newblock Stability of polyatomic molecules in degenerate electronic states - i
  - orbital degeneracy.
\newblock {\em Proceedings of the Royal Society of London. Series A -
  Mathematical and Physical Sciences}, 161(905):220--235, 1937.

\bibitem{Wu2016}
Chuanjian Wu, Zhong Yu, Ke~Sun, Jinlan Nie, Rongdi Guo, Hai Liu, Xiaona Jiang,
  and Zhongwen Lan.
\newblock Calculation of exchange integrals and curie temperature for
  \text{La}-substituted barium hexaferrites.
\newblock {\em Scientific Reports}, 6(1):36200, Oct 2016.

\bibitem{NovakPRB05}
P.~Nov\'ak and J.~Rusz.
\newblock Exchange interactions in barium hexaferrite.
\newblock {\em Phys. Rev. B}, 71:184433, May 2005.

\bibitem{ZiJMM08}
Z.F. Zi, Y.P. Sun, X.B. Zhu, Z.R. Yang, J.M. dai, and W.H. Song.
\newblock Structural and magnetic properties of srfe$_{12}$o$_{19}$ hexaferrite
  synthesized by a modified chemical co-precipitation method.
\newblock {\em Journal of Magnetism and Magnetic Materials}, 320(21):2746 --
  2751, 2008.

\bibitem{Ralph}
Ralph Skomski.
\newblock {\em Simple Models of Magnetism}.
\newblock Oxford University Press, Oxford University Press Inc., New York,
  2008.

\bibitem{JahnPSS69}
L.~Jahn and H.~G. Müller.
\newblock The coercivity of hard ferrite single crystals.
\newblock {\em physica status solidi (b)}, 35(2):723--730, 1969.

\bibitem{ShirkJAP69}
B.~T. Shirk and W.~R. Buessem.
\newblock Temperature dependence of ms and k1 of
  $\text{BaFe}_{12}\text{O}_{19}$ and \text{SrFe}$_{12}$\text{O}$_{19}$ single
  crystals.
\newblock {\em Journal of Applied Physics}, 40(3):1294--1296, 1969.

\end{thebibliography}
\bibliographystyle{unsrt}

\end{document}